\documentclass{aa}

\usepackage[]{graphics}
\usepackage[]{psfig}

\begin{document}

\thesaurus{08.01.1;08.03.2;08.12.1;10.07.3 $\omega$ Cen; 10.07.3 M22}

\title{Revised Str\"omgren metallicity calibration for red giants
\thanks{Based on data collected at the European Southern Observatory,
         La Silla, Chile}}

\author {Michael Hilker
} 

\offprints {M.~Hilker}
\mail{mhilker@astro.puc.cl}

\institute{
Departamento de Astronom\'\i a y Astrof\'\i sica, P.~Universidad Cat\'olica,
Casilla 104, Santiago 22, Chile
}

\date {Received --- / Accepted ---}

\titlerunning{Revised Str\"omgren metallicity calibration for red giants} 
\authorrunning{M.~Hilker}
\maketitle

\begin{abstract}

A new calibration of the Str\"omgren $(b-y),m_1$ diagram in terms of iron 
abundance of red giants is presented. This calibration is 
based on a homogeneous sample of giants in the globular clusters $\omega$ 
Centauri, M22, and M55 as well as field
giants from the list of Anthony-Twarog \& Twarog (\cite{anth98}). Towards high
metallicities,
the new calibration is connected to a previous calibration by Grebel \&
Richtler (\cite{greb92}), which was unsatisfactory for iron abudances lower 
than $-$1.0 dex. The revised calibration is valid for CN-weak/normal 
red giants in the abundance range of 
$-2.0 <$[Fe/H]$< 0.0$ dex, and a color range of $0.5 < (b-y) < 1.1$ mag.

As shown for red giants in $\omega$ Centauri, CN-weak stars with Str\"omgren
metallicities higher than $-$1.0 dex cannot be distinguished in the $(b-y),m_1$
diagram from stars with lower iron abundances but higher CN band strengths. 


\end{abstract}


\section{Introduction}

Str\"omgren photometry has been proven to be a
reliable metallicity indicator for globular cluster giants and
subgiants (e.g. Richtler \cite{rich89}, Grebel \& Richtler \cite{greb92} and 
references therein). 
The location of late type (G and K) giants in the
Str\"omgren $(b-y),m_1$ diagram is correlated with their metallicities,
especially with their iron and CN abundances. Whereas the color $(b-y)$ is not
sensitive to metallicity, the Str\"omgren $v$ filter includes several iron 
absorption lines as well as the CN band at 4215\AA, and therefore $m_1 =
(v-b) - (b-y)$ is a metallicity sensitive index (e.g. Bell \& Gustafsson 
\cite{bell}). Within a certain color range, $0.5 < (b-y) < 1.1$ mag,
the loci of constant iron abundance of giants and supergiants can be 
approximated 
by straight lines. This is valid for CN-``normal'' ($=$ CN-weak) stars.
CN-strong stars, due to their higher absorption in the $v$ filter, scatter to
higher $m_1$ values and therefore mimic a higher Str\"omgren metallicity
than their actual iron abundance would correspond to. If a cluster is known
to have only small star-to-star variations in its iron abundance, the scatter 
to higher $m_1$ values 
can be used to uncover CN-rich member stars. For giants redder than
$(b-y) = 1.1$ mag the calibration breaks down due to TiO and MgH absorption 
in the $y$ band.

Until now the calibration for Str\"omgren metallicities is based on
the spectroscopically determined iron abundance of relatively metal-rich stars,
and only few metal poorer stars (Grebel \& Richtler \cite{greb92}).
This calibration worked well
for the determination of cluster abundances in the Magellanic Clouds
(Hilker et al. \cite{hilk95a}, \cite{hilk95b}; Dirsch et al. \cite{dirs}),
but failed to reproduce the slope of constant metallicity lines in the
$(b-y),m_1$ diagram for the new data of metal poor globular clusters. 
Other previous metallicity calibrations in the Str\"omgren $(b-y),m_1$ plane
have been published for the regime of F and G dwarfs (Schuster \& Nissen 
\cite{schu}) and F dwarfs and giants in the small color range of $0.22 < (b-y) <
0.38$ (Malyuto \cite{maly}). 

In the meantime, large samples of homogeneous spectroscopic measurements of
cluster as well as field giants have become available. Furthermore, 
CCD Str\"omgren photometry, in particular in globular clusters, is becoming
popular (e.g. Grundahl et al. \cite{grun99}).
Therefore, a revised, more accurate calibration of the metallicity sensitive
Str\"om- gren indices is needed.
In this paper, the Str\"omgren metallicity calibration is extended to 
more metal-poor stars ([Fe/H] $>-2.0$ dex). The sample of stars used for the
new calibration contains red giants of $\omega$ Centauri, M55, and M22, and 
field
giants from a homogenized sample of Anthony-Twarog \& Twarog (\cite{anth98}).
$\omega$ Cen is known for its star-to-star variations in CN as well as iron 
abundances (e.g. Vanture et al. \cite{vant}). For 40 giants in $\omega$ Cen, 
Norris \& Da Costa (\cite{norr95})
measured accurate abundances from high resolution spectroscopy. M22 also shows
CN abundance variations (e.g. Norris \& Freeman \cite{norr82}, \cite{norr83}).
Both clusters have been used to study the influence of the CN band strengths 
on the Str\"omgren metallicity.

\begin{figure}
\psfig{figure=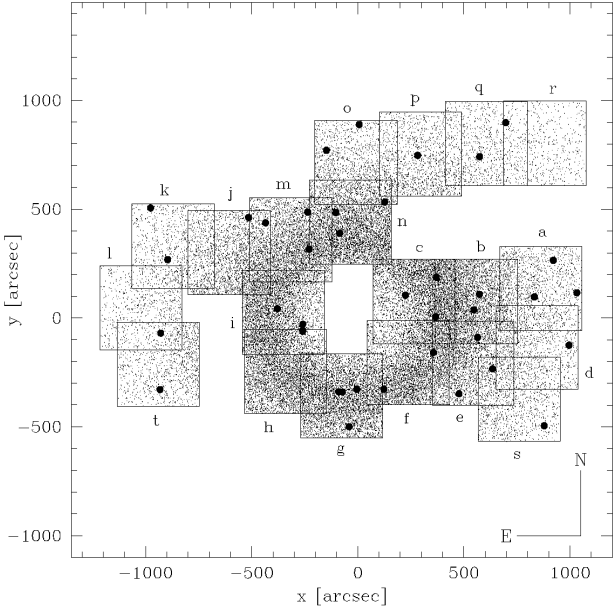,height=8.6cm,width=8.6cm
,bbllx=9mm,bblly=65mm,bburx=195mm,bbury=246mm}
\vspace{0.4cm}
\caption{Position plot of the observed fields in $\omega$ Cen. All stars
with a $V$ magnitude brighter than 19.0 mag and a photometric error less than
0.1 mag have been plotted. Filled circles indicate the position of the red 
giants with known spectroscopic abundances
}
\end{figure}

\begin{table}
\caption[]{\label{log} Position log of $\omega$ Cen fields}
\begin{flushleft}
{\small
\begin{tabular}{cccccc}
\hline\noalign{\smallskip}
Id. & $\alpha_{2000}$ & $\delta_{2000}$ & Id. & $\alpha_{2000}$ &
$\delta_{2000}$ \\
\noalign{\smallskip}
\hline\noalign{\smallskip}
a & 13:25:30.8 & $-$47:26:46 & k & 13:28:12.9 & $-$47:23:36 \\
b & 13:25:59.9 & $-$47:17:46 & l & 13:28:27.0 & $-$47:28:15 \\
c & 13:26:28.0 & $-$47:27:46 & m & 13:27:18.0 & $-$47:23:05 \\
d & 13:25:30.8 & $-$47:31:16 & n & 13:26:50.0 & $-$47:21:47 \\
e & 13:26:00.0 & $-$47:32:27 & o & 13:26:47.0 & $-$47:17:17 \\
f & 13:26:29.0 & $-$47:32:27 & p & 13:26:16.9 & $-$47:16:38 \\
g & 13:26:59.0 & $-$47:34:56 & q & 13:25:46.9 & $-$47:15:47 \\
h & 13:27:24.0 & $-$47:33:06 & r & 13:25:29.9 & $-$47:15:47 \\
i & 13:27:24.0 & $-$47:28:34 & s & 13:25:30.8 & $-$47:35:07 \\
j & 13:27:48.0 & $-$47:24:05 & t & 13:28:15.9 & $-$47:32:26 \\
\noalign{\smallskip}
\hline
\end{tabular}
}
\end{flushleft}
\end{table}

\section{Observations and data reduction: $\omega$ Cen, M55, and M22}

The observations were performed during the nights  21-24 April 1995
with the Danish 1.54m telescope at ESO/La Silla. The CCD in use
was a Tektronix chip with 1024$\times$1024 pixels.
The $f$/8.5 beam of the telescope provides a scale of $15\farcs7$/mm,
and with a pixel size of 24 $\mu$m the total field is $6\farcm3 \times
6\farcm3$.
The observations and data reduction of the M22 and M55 images have been
presented by Richter et al. (\cite{richp}).
In $\omega$ Cen, 20 fields were observed during the third
night of the run through the Str\"omgren $ybv$ filters (Danish set of imaging
filters). Table \ref{log}
gives a log of the field positions. In Fig.~1, all fields are plotted in
a coordinate system centered on $\omega$ Cen. The exposure times
were 70, 120, and 240 seconds for $y$, $b$ and $v$, respectively. All nights 
had photometric conditions, and the seeing, measured from the
FWHM of stellar images, was in the range $1\farcs2 - 1\farcs5$.

\begin{figure}
\psfig{figure=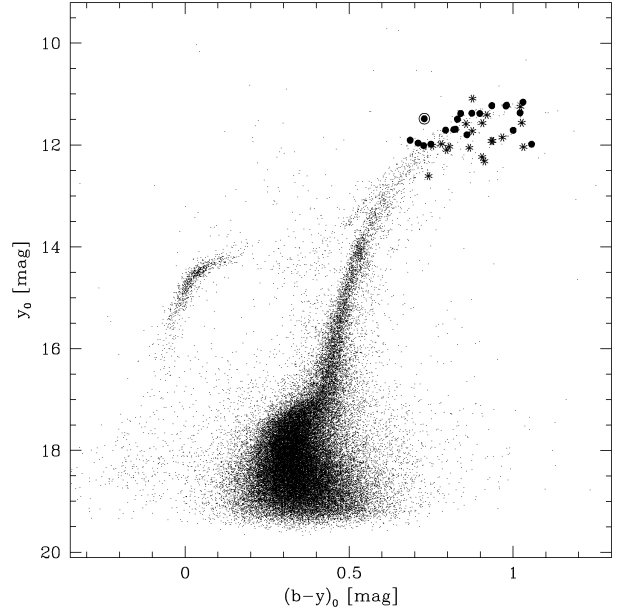,height=8.6cm,width=8.6cm
,bbllx=9mm,bblly=65mm,bburx=195mm,bbury=246mm}
\vspace{0.4cm}
\caption{Color magnitude diagram of all stars in the observed fields (see
Fig.~1) with photometric errors in all colors smaller than 0.1 mag. Filled 
circles
indicate red giants from the spectroscopic sample with normal CN abundances,
whereas red giants marked with asterisks are known to be CN-strong. The 
encircled dot is the star ROA 74 which deviates most from the metallicity 
calibration
}
\end{figure}

The CCD frames were processed with standard IRAF routines. Instrumental
magnitudes were derived using DAO- PHOT II (Stetson \cite{stet87}, 
\cite{stet92}). For the
comparison with the standard stars, aperture--PSF shifts have been determined
in all fields. The remaining uncertainty of this shift is in the order of 0.01
mag in all filters. The corrected magnitudes of the stars belonging to
overlapping areas of two adjacent fields agree very well and have been avaraged
for the final photometry file. The calibration equations and coefficients
for the third night are given in Richter et al. (\cite{richp}).

\begin{table*}
\caption{Photometric and spectroscopic data for red giants in $\omega$ Cen.
All columns are explained in Sect.~3.1. A reddening of $E_{B-V} = 0.11$ mag
has been applied}
\begin{flushleft}
\begin{tabular}{rrrcccccccccc}
\noalign{\smallskip}
\hline
ROA & $\Delta x$ & $\Delta y$ & $V_0$ & $(b-y)_0$ & $m_{1,0}$ & [Fe/H]$_{\rm
ph}$ & [Fe/H]$_{\rm sp}$ & CO & CN & C4142 & S3839 & CB \\
\noalign{\smallskip}
\hline
\noalign{\smallskip}
  40 &  564.0 &  $-$89.0 & 11.088 & 0.876 & 0.484 & $-$1.21 & $-$1.69 & $\circ$   &     ...   &  ... &  0.46 & 0.34 \\
  42 & $-$261.0 &  $-$56.0 & 11.409 & 0.919 & 0.506 & $-$1.27 & $-$1.69 &    ...    &  $\circ$  & 0.32 &  0.50 &  ... \\
  43 &  696.5 &  897.9 & 11.249 & 1.022 & 0.671 & $-$1.01 & $-$1.47 & $\circ$   &     ...   & 0.34 &  0.42 & 0.31 \\
  46 & $-$103.1 &  485.2 & 11.232 & 0.978 & 0.472 & $-$1.57 & $-$1.67 & $\bullet$ & $\bullet$ & 0.19 &   ... &  ... \\
  48 & $-$230.1 &  316.8 & 11.219 & 0.980 & 0.422 & $-$1.75 & $-$1.76 & $\bullet$ & $\bullet$ &  ..  &   ... & 0.19 \\
  53 &  878.1 & $-$494.7 & 11.157 & 1.030 & 0.561 & $-$1.40 & $-$1.67 & $\bullet$ & $\bullet$ & 0.24 &  0.16 & 0.22 \\
  58 & $-$236.0 &  486.5 & 11.376 & 0.874 & 0.396 & $-$1.55 & $-$1.73 & $\bullet$ & $\bullet$ & 0.21 &  0.16 & 0.22 \\
  65 &  369.5 &  188.0 & 11.225 & 0.935 & 0.369 & $-$1.82 & $-$1.72 &    ...    & $\bullet$ &  ... &  0.03 &  ... \\
  74 &  126.5 &  533.2 & 11.483 & 0.729 & 0.498 & $-$0.52 & $-$1.80 & $\bullet$ & $\bullet$ & 0.17 &  0.16 & 0.21 \\
  84 &  $-$85.7 &  389.2 & 11.565 & 1.026 & 0.708 & $-$0.90 & $-$1.36 & $\circ$   &     ...   & 0.30 &  0.17 & 0.30 \\
  91 &    6.4 &  889.4 & 11.491 & 0.830 & 0.329 & $-$1.69 & $-$1.73 & $\bullet$ & $\bullet$ &  ... &  0.07 & 0.16 \\
  94 &  572.8 &  108.9 & 11.379 & 0.840 & 0.330 & $-$1.72 & $-$1.78 &    ...    & $\bullet$ & 0.15 &  0.03 &  ... \\
 100 & $-$259.6 &  $-$28.3 & 11.568 & 0.906 & 0.762 & $-$0.25 & $-$1.49 &    ...    &  $\circ$  & 0.51 &  0.58 &  ... \\
 102 & $-$929.1 &  $-$69.2 & 11.380 & 0.898 & 0.358 & $-$1.77 & $-$1.80 & $\bullet$ & $\bullet$ &  ... &  0.22 &  ... \\
 132 &  $-$90.7 & $-$339.1 & 11.367 & 1.021 & 0.634 & $-$1.13 & $-$1.37 & $\circ$   &     ...   & 0.22 &   ... & 0.28 \\
 139 & $-$435.2 &  437.5 & 11.585 & 0.856 & 0.618 & $-$0.60 & $-$1.46 & $\circ$   &  $\circ$  & 0.39 &   ... & 0.52 \\
 144 & $-$260.2 &  $-$61.0 & 11.980 & 0.781 & 0.473 & $-$0.89 & $-$1.66 &    ...    &  $\circ$  & 0.33 &   ... &  ... \\
 155 &  994.8 & $-$124.6 & 11.689 & 0.824 & 0.421 & $-$1.28 & $-$1.64 & $\bullet$ & $\bullet$ & 0.17 &   ... & 0.26 \\
 159 &  921.0 &  265.2 & 11.708 & 0.794 & 0.323 & $-$1.60 & $-$1.72 & $\bullet$ & $\bullet$ & 0.16 &   ... & 0.14 \\
 161 & $-$515.7 &  461.7 & 11.697 & 0.819 & 0.396 & $-$1.37 & $-$1.67 & $\bullet$ & $\bullet$ & 0.14 &   ... & 0.26 \\
 162 & $-$932.5 & $-$328.7 & 11.851 & 0.967 & 0.965 & $+$0.21 & $-$1.10 & $\bullet$ &  $\circ$  & 0.54 &   ... & 0.49 \\
 171 &  634.0 & $-$233.5 & 11.727 & 0.875 & 0.548 & $-$0.94 & $-$1.43 & $\circ$   & $\bullet$ & 0.18 &   ... & 0.31 \\
 179 & $-$379.6 &   43.2 & 11.710 & 1.000 & 0.710 & $-$0.81 & $-$1.10 & $\circ$   &     ...   & 0.18 &   ... & 0.29 \\
 182 &  $-$72.8 & $-$340.3 & 11.800 & 0.859 & 0.515 & $-$1.03 & $-$1.46 &    ...    &     ...   & 0.23 &   ... &  ... \\
 201 &  224.4 &  105.0 & 11.984 & 1.056 & 0.692 & $-$1.05 & $-$0.85 &    ...    &     ...   &  ... &   ... & 0.28 \\
 213 &  281.7 &  747.5 & 11.904 & 0.686 & 0.211 & $-$1.76 & $-$1.83 & $\bullet$ & $\bullet$ & 0.18 &   ... & 0.06 \\
 219 & $-$896.1 &  269.1 & 11.922 & 0.934 & 0.639 & $-$0.83 & $-$1.25 & $\circ$   &     ...   & 0.20 &   ... & 0.32 \\
 231 &  355.7 & $-$158.4 & 11.919 & 0.937 & 0.818 & $-$0.18 & $-$1.10 & $\bullet$ &  $\circ$  & 0.40 &   ... & 0.33 \\
 234 & $-$148.1 &  771.0 & 11.962 & 0.710 & 0.222 & $-$1.79 & $-$1.78 & $\bullet$ & $\bullet$ & 0.11 & $-$0.33 & 0.06 \\
 248 &  122.4 & $-$326.7 & 12.039 & 1.031 & 0.968 & $-$0.06 & $-$0.78 & $\bullet$ &     ...   & 0.42 &  0.51 & 0.39 \\
 252 &  573.5 &  741.2 & 11.983 & 0.749 & 0.250 & $-$1.79 & $-$1.74 &    ...    & $\bullet$ & 0.10 &  0.10 &  ... \\
 253 & 1031.6 &  116.0 & 12.014 & 0.753 & 0.599 & $-$0.18 & $-$1.39 & $\bullet$ &  $\circ$  & 0.42 &  1.00 & 0.57 \\
 256 &  546.2 &   37.4 & 12.012 & 0.727 & 0.301 & $-$1.47 & $-$1.58 & $\bullet$ & $\bullet$ & 0.14 &  0.27 &  ... \\
 270 &  476.6 & $-$347.5 & 12.055 & 0.866 & 0.655 & $-$0.49 & $-$1.22 & $\circ$   &     ...   &  ... &  0.57 & 0.39 \\
 279 & $-$976.9 &  505.9 & 12.035 & 0.805 & 0.643 & $-$0.26 & $-$1.69 &    ...    &    CH     & 0.35 &  1.15 & 0.59 \\
 287 &  832.0 &   97.4 & 12.098 & 0.797 & 0.546 & $-$0.64 & $-$1.43 & $\circ$   &  $\circ$  & 0.23 &  0.58 & 0.42 \\
 357 &  366.3 &    5.9 & 12.236 & 0.905 & 0.877 & $+$0.19 & $-$0.85 & $\bullet$ &     ...   &  ... &   ... & 0.41 \\
 371 &  $-$41.0 & $-$499.5 & 12.320 & 0.912 & 0.847 & $+$0.04 & $-$0.79 & $\bullet$ &     ...   & 0.46 &  0.54 & 0.39 \\
 480 &   $-$5.2 & $-$327.8 & 12.611 & 0.742 & 0.797 & $+$0.82 & $-$0.95 & $\bullet$ &  $\circ$  &  ... &  1.32 & 0.65 \\
\noalign{\smallskip}
\hline
\end{tabular}
\end{flushleft}
\end{table*}

After the photometric reduction and calibration
of the magnitudes, the average photometric errors for the red giants used in the
metallicity calibration are 0.015 mag for $V$, 0.016 mag for $(b-y)$ and 0.024
mag for $m_1$.

%
%

\section{Sample of red giant stars}

This section presents the sample of red giants that has been used for our
new metallicity calibration. The spectroscopically determined iron abundances
of the different authors are all consistent with the Zinn \& West (\cite{zinn})
abundance scale. All Str\"omgren colors refer to the photometric system 
defined by Olsen (\cite{olse93}). The photometric colors of the previous
calibration (Grebel \& Richtler \cite{greb92}) are based on the system of
primary standards by Bond (\cite{bond}) and Olsen (\cite{olse83}, \cite{olse84})
and have been corrected to the Olsen (1993) system according to the 
transformations given by Olsen (\cite{olse95}). In the same way the $m_1$ color
of the field stars sample of Anthony-Twarog \& Twarog (\cite{anth98}) has been
corrected according to Olsen's transformations. The $V$ and $(b-y)$ colors in 
this sample are on the system of Olsen (1993).

For our calibration, we used 12 E region stars from J{\o}nch-S{\o}rensen
(\cite{jonc93}) and 5 stars from his 1994 list (\cite{jonc94}), namely E3-33, 
E4-37, E4-108, E5-32, E5-48, E5-56, E6-48, E6-98, E7-64, E8-39, E8-47, E8-48, 
and F4-2, F5-2, F5-3, F6-1, F6-3. Their colors are consistent with the Olsen
(1993) photometric system. The standard stars are uniformly distributed over 
the color range $0.2 < (b-y) < 1.3$ mag which has been used for our 
metallicity calibration.

\begin{figure}
\psfig{figure=mhilker.f3.ps,height=11.3cm,width=8.6cm
,bbllx=9mm,bblly=65mm,bburx=155mm,bbury=246mm}
\vspace{0.4cm}
\caption{The cyanogen indices C4142, S3839, and CB of red giants in
$\omega$ Cen are plotted versus their spectroscopically determined iron
abundances [Fe/H]$_{\rm sp}$ (left panels) and versus the difference between
[Fe/H]$_{\rm ph}$ and [Fe/H]$_{\rm sp}$.
The horizontal dotted lines give the separation criteria between CN-strong
(asterisks) and CN-weak stars (filled circles). Clearly, most CN-strong stars
lie beyond the $\pm0.25$ dex error range of the metallicity calibration
(vertical dotted lines in the right panels).
The encircled dot is star ROA 74 which deviates most from the metallicity
calibration. The open triangles in the middle panels indicate the location
of M22 giants which also follow the general trend seen in $\omega$ Cen
}
\end{figure}

\begin{figure}
\psfig{figure=mhilker.f4.ps,height=8.6cm,width=8.6cm
,bbllx=9mm,bblly=65mm,bburx=195mm,bbury=246mm}
\vspace{0.4cm}
\caption{Two-color diagram for the spectroscopic sample of red giants in
$\omega$ Cen. As in Fig.~3 CN-rich stars are marked with asterisks,
CN-weak with filled circles. Star ROA 74 is encircled. The error bars are
calculated from the photometric error, the calibration error, and the error
for aperture-PSF shifts. The lines indicate the two metallicity calibrations 
(solid: Eq.~(1); dashed: Eq.~(3)) between $-$2.0 and 0.0 dex
}
\end{figure}

\subsection{$\omega$ Centauri}

The fields selected and observed in $\omega$ Cen were chosen to cover the
40 red giants for which accurate abundances from high
resolution spectroscopy have been published (Norris \& Da Costa \cite{norr95}).
All red giants in this sample are amongst the brightest stars of the red giant
branch (RGB) of $\omega$ Cen (see Fig.~2). In their iron abundance, they
span a range between $-$1.8 and $-$0.8 dex. Of these stars, 11 are known to be
CN-strong, relatively to the average CN band strengths of the whole sample.

\begin{table*}
\caption{Photometric and spectroscopic data for red giants in M22. All columns
are explained in Sect.~3.3. A reddening of $E_{B-V} = 0.42$ mag has been
applied}
\begin{flushleft}
\begin{minipage}{13cm}
\begin{tabular}{lcccccccccl}
\noalign{\smallskip}
\hline
Id$_{Arp}$ & $V_0$ & $(b-y)_0$ & $m_{1,0}$ & [Fe/H]$_{\rm
ph}$ & [Fe/H]$_{\rm sp}$ & Ref.$^a$ & CN & S3839 & RGB \\
\noalign{\smallskip}
\hline
\noalign{\smallskip}
I12   & 10.338 & 0.705 & 0.261 & $-$1.58 & $-$1.72 & BW  & $\bullet$ & 0.16 & blue\\
I53   & 11.343 & 0.616 & 0.266 & $-$1.16 & $-$1.63 & LBC & $\circ$   & 0.54 & red \\
I80   & 11.204 & 0.606 & 0.339 & $-$0.67 & $-$1.42 & LBC & $\circ$   & 0.92 & blue\\
I85   & 11.136 & 0.573 & 0.153 & $-$1.63 & $-$1.56 & LBC & $\bullet$ & 0.11 & blue\\
I86   & 10.982 & 0.675 & 0.191 & $-$1.82 & $-$1.65 & BW  & $\bullet$ & 0.05 & red \\
I92   & 10.220 & 0.756 & 0.320 & $-$1.48 & $-$1.55 & CG  & $\bullet$ & 0.08 & red \\
I108  & 11.480 & 0.505 & 0.106 & $-$1.60 & $-$1.41 & LBC & $\bullet$ & 0.01 & blue\\
I116  & 11.481 & 0.540 & 0.111 & $-$1.74 & $-$1.56 & LBC & $\circ$ & 0.35 & blue\\
II92  & 11.253 & 0.638 & 0.330 & $-$0.91 & $-$1.51 & BW  & $\circ$   & 0.76 & red \\
III3  & 9.828 & 0.881 & 0.651 & $-$0.57 & $-$1.50 & BW  & $\circ$   & 0.50 & blue\\
III12 & 10.201 & 0.812 & 0.540 & $-$0.73 & $-$1.49 & CG  & $\circ$   & 0.58 & red \\
III25 & 11.338 & 0.556 & 0.114 & $-$1.80 & $-$1.49 & LBC & $\circ$   & 0.41 & blue\\
III35 & 11.051 & 0.636 & 0.186 & $-$1.70 & $-$1.72 & LBC & $\bullet$ & 0.11 & blue\\
III52 & 10.254 & 0.787 & 0.637 & $-$0.19 & $-$1.46 & BW  & $\circ$   & 0.55 & red \\
IV20  & 11.739 & 0.727 & 0.424 & $-$0.87 & $-$1.53 & BW  & $\circ$   & 0.72 & red \\
IV24  & 11.282 & 0.618 & 0.369 & $-$0.57 & $-$1.57 & LBC & $\circ$   & 1.05 & red \\
\noalign{\smallskip}
\hline
\end{tabular}
{\footnotesize $^a$ BW = Brown \& Wallerstein \cite{brow92}, LBC = Lehnert et
al. \cite{lehn}, CG = Carretta \& Gratton \cite{care}}
\end{minipage}
\end{flushleft}
\end{table*}

In Table 2 the photometric and spectroscopic data that are relevant for our
analysis are summarised. The numbering (column 1) is from Woolley
et al. (\cite{wool}). Columns 2 and 3 give the coordinates in x,y distances
(in arcsec) to the center of $\omega$ Cen, as shown in Fig.\,1 (center:
$\alpha_{2000} = 13^h26^m46.8^s$ and $\delta_{2000} =
-47\deg28\arcmin47\arcsec$, Lyng{\aa} \cite{lyng96}). In columns 4, 5, and 6
the reddening corrected Str\"omgren values $V_0$, $(b-y)_0$, and $m_{1,0}$ are
given. A reddening of $E_{B-V} = 0.11$ (Zinn \cite{zinn85}, Webbink
\cite{webb}) has been adopted. This corresponds
to $E_{b-y} = 0.08$ and $E_{m_1} = -0.02$, using the relations $E_{b-y} = 
0.7E_{B-V}$ and $E_{m_1} = -0.3E_{b-y}$ (Crawford \& Barnes \cite{craw}).
In column 7 we list the metallicity [Fe/H]$_{\rm ph}$ as derived from our
new calibration (see Sect.~4, Eq.~(1)). The spectroscopic parameters in the
columns
8 to 13 are taken from the compilation of Norris \& Da Costa (\cite{norr95}).
[Fe/H]$_{\rm sp}$ is the iron abundance determined by them. ``CO'' indicates
CO-strong (open circles) and CO-weak (filled circles) stars as determined from
the strength of the CO molecule by Persson et al. (\cite{pers}). ``CN'' is the
relative strength of the CN bands, open circles indicate CN-strong stars, filled
circles CN-weak stars. Note that CN-weak can be understood as CN-normal when
compared to the average CN band strengths in other clusters.
The star ROA 279 is known to be a CH star.
In the columns 11, 12 and 13, the cyanogen indices
C4142 and S3839, and the mean violet index CB are given as determined by
different authors (see Norris \& Da Costa, \cite{norr95}, for references).
C4142 is an index of the DDO filter system (McClure \cite{mccl}) that compares
the flux in the filter 41 ($\lambda_0 = 4166$\AA, $\Delta\lambda = 83$\AA) which
includes a violet cyanogen-band absorption with that of filter 42 ($\lambda_0
= 4257$\AA, $\Delta\lambda = 73$\AA). S3839 compares the intensity in the violet
CN band at $\approx 3883$\AA with the nearby continuum (Norris et al.
\cite{norr81}). CB is the mean $3883\AA$ CN band strength as determined from two
independent measurements (Cohen \& Bell \cite{cohe86}).

\subsection{The influence of CN band strengths}

Since the CN bands fall in the range of the Str\"omgren $v$ filter, and
therefore influences the $m_1$ index, CN-strong stars have to be excluded
from our metallicity calibration. Only CN-weak stars will not disturb the
transformation of Str\"omgren colors into an iron abundance.
In Fig.~3 we show the 3 cyanogen indices C4142, S3839, and CB as a function of
on the iron abundance (left panels) and as a function of the difference between
the photometrically and spectroscopically determined iron abundances,
$\Delta$[Fe/H] = [Fe/H]$_{\rm ph} - $[Fe/H]$_{\rm sp}$ (right panels).
All stars that have a C4142, S3839, or CB value higher than 0.27, 0.33, and
0.29 respectively, have been assigned to be CN-strong stars, and are marked with
asterisks. The separation of CN-strong and CN-weak stars can clearly be seen
for the C4142 and S3839 index, and the separation values are chosen to be
located in the middle of the gap. For the CB index, the separation at 0.29 is
motivated by the fact that all stars with CB $>$ 0.29 significantly
deviate in [Fe/H]$_{\rm ph}$ from [Fe/H]$_{\rm sp}$, whereas most stars
with CB $<$ 0.29 have [Fe/H]$_{\rm ph}$ values within the $\pm0.25$ dex
error range of [Fe/H]$_{\rm sp}$. This also is true for the C4142 and S3839
index (see right panels in Fig.~3). Outliers with a normal CN band strength,
but a large $\Delta$[Fe/H], are the stars ROA 53, 74, 155, 161, 179,
and 182. Whereas ROA 53 and 182 have a quite high C4142 value, and ROA 155,
161, and 179 a high CB index, no obvious explanation can be found for ROA 74.
Having moderately low values of all CN indices, the determined Str\"omgren
metallicity of ROA 74 is more than 1 dex too high compared to its
spectroscopic metallicity. However, this star is not
located on the average RGB of $\omega$ Cen (see Fig.~3, encircled dot),
but is about 0.1 mag bluer. Examination of the image and photometry of this star
reveals that its blue color is due to an overlapping blue star that
could not be separated by PSF fitting.

In general, 
$\Delta$[Fe/H] is correlated to the CN band strengths. The more CN-rich the
star, the higher $\Delta$[Fe/H]. This might be used to determine the iron
abundance of a red giant if its Str\"omgren colors and one of the CN indices
are known, or conversely to detect CN-rich stars when their iron
abundances are known.

In Fig.~4 the distribution of $\omega$ Cen red giants in the two-color
diagram $m_{1,0}$ versus $(b-y)_0$ is shown. The symbols are the same as in
Fig.~3.
The lines of constant metallicity between $-$2.0 and 0.0 dex according to
our new metallicity calibrations (Sect.~4) are indicated.
The CN-weak stars are located in a metallicity range between $-$2.0 and $-$1.0
dex, the known iron abundance range for $\omega$ Cen.
In contrast, CN-rich stars, although having the same spectroscopic iron
abundances as the CN-weak stars, range between $-$1.0 and 0.5 dex.
Around a metallicity of $-$1.0 dex, CN-weak stars with this iron
abundance can not be distinguished from stars with lower iron abundances
(between $-$1.7 and $-$1.4 dex), but higher CN band strengths.

For our new metallicity calibration, only CN-weak red giants of the 
spectroscopic sample (filled circles in
Figs. 2, 3 and 4) have been selected for the Str\"omgren metallicity
calibration, and CN-rich stars (asterisks) have been excluded.

\begin{table}
\caption{Photometric and spectroscopic data for red giants of the 
Anthony-Twarog \& Twarog (\cite{anth98}) sample. The $m_{1,0}$ index is 
transformed to the system of Olsen (\cite{olse93}, see Olsen \cite{olse95}).
[Fe/H]$_{\rm ph}$ is the metallicity as determined from our new calibration 
(Eq.~(1))}
\begin{flushleft}
{\scriptsize
\begin{tabular}{lrcccc}
\noalign{\smallskip}
\hline
Name & $V_0$ & $(b-y)_0$ & $m_{1,0}$ & [Fe/H]$_{\rm sp}$ & [Fe/H]$_{\rm ph}$ \\
\noalign{\smallskip}
\hline
\noalign{\smallskip}
HD 85773     &  9.326 & 0.773 & 0.163 & $-$2.28 & $-$2.26 \\
HD 165195    &  6.891 & 0.828 & 0.173 & $-$2.25 & $-$2.34 \\
HD 23798     &  8.307 & 0.742 & 0.181 & $-$2.22 & $-$2.09 \\
HD 216143    &  7.767 & 0.681 & 0.148 & $-$2.13 & $-$2.07 \\
HD 103545    &  9.433 & 0.589 & 0.089 & $-$2.09 & $-$2.09 \\
HD 36702     &  8.329 & 0.823 & 0.216 & $-$2.03 & $-$2.15 \\
BD $-$14 5890 & 10.130 & 0.535 & 0.091 & $-$1.99 & $-$1.85 \\
HD 222434    &  8.793 & 0.708 & 0.221 & $-$1.94 & $-$1.78 \\
HD 104893    &  9.028 & 0.780 & 0.221 & $-$1.92 & $-$2.01 \\
HD 3008      &  9.393 & 0.838 & 0.295 & $-$1.90 & $-$1.85 \\
HD 136316    &  7.339 & 0.751 & 0.230 & $-$1.85 & $-$1.88 \\
HD 204543    &  8.203 & 0.615 & 0.151 & $-$1.78 & $-$1.81 \\
BD +01 2916 &  9.613 & 0.881 & 0.306 & $-$1.78 & $-$1.93 \\
HD 118055    &  8.716 & 0.806 & 0.290 & $-$1.75 & $-$1.78 \\
HD 122956    &  7.054 & 0.626 & 0.190 & $-$1.72 & $-$1.63 \\
HD 21581     &  8.557 & 0.530 & 0.122 & $-$1.72 & $-$1.60 \\
HD 26297     &  7.477 & 0.740 & 0.232 & $-$1.69 & $-$1.84 \\
HD 126238    &  7.544 & 0.525 & 0.128 & $-$1.69 & $-$1.53 \\
HD 187111    &  7.380 & 0.757 & 0.228 & $-$1.69 & $-$1.91 \\
HD 220838    &  9.359 & 0.764 & 0.265 & $-$1.68 & $-$1.76 \\
HD 8724      &  8.196 & 0.659 & 0.155 & $-$1.64 & $-$1.96 \\
HD 141531    &  9.070 & 0.754 & 0.270 & $-$1.61 & $-$1.70 \\
HD 206739    &  8.460 & 0.603 & 0.180 & $-$1.60 & $-$1.59 \\
HD 220662    & 10.086 & 0.686 & 0.189 & $-$1.60 & $-$1.87 \\
HD 83212     &  8.251 & 0.678 & 0.252 & $-$1.48 & $-$1.51 \\
HD 37828     &  6.611 & 0.683 & 0.325 & $-$1.32 & $-$1.14 \\
HD 111721    &  7.946 & 0.511 & 0.155 & $-$1.31 & $-$1.25 \\
HD 128188    &  9.752 & 0.592 & 0.204 & $-$1.29 & $-$1.38 \\
HD 99978     &  8.547 & 0.575 & 0.271 & $-$1.03 & $-$0.86 \\
HD 171496    &  7.693 & 0.575 & 0.182 & $-$1.03 & $-$1.43 \\
HD 7595      &  9.691 & 0.754 & 0.460 & $-$0.85 & $-$0.81 \\
HD 24616     &  6.691 & 0.505 & 0.226 & $-$0.82 & $-$0.67 \\
HD 81223     &  8.245 & 0.584 & 0.272 & $-$0.79 & $-$0.91 \\
HD 11722     &  8.936 & 0.523 & 0.217 & $-$0.76 & $-$0.87 \\
BD $-$18 2065 &  9.567 & 0.573 & 0.270 & $-$0.67 & $-$0.85 \\
CP $-$57 0680 &  9.268 & 0.579 & 0.265 & $-$0.60 & $-$0.92 \\
HD 35179     &  9.348 & 0.576 & 0.287 & $-$0.59 & $-$0.76 \\
HD 81713     &  8.851 & 0.584 & 0.284 & $-$0.47 & $-$0.83 \\
\noalign{\smallskip}
\hline
\end{tabular}
}
\end{flushleft}
\end{table}

\subsection{M22 \& M55}

The globular clusters M22 and M55 were observed in the same observing run 
as $\omega$ Cen. The results are presented in Richter et al. (\cite{richp}).
In M22, 16 red giants in our observed fields have known iron abundances and CN 
band strengths (Carretta \& Gratton \cite{care}, Brown \& Wallerstein 
\cite{brow92}, Lehnert et al. \cite{lehn}, Norris \& Freeman \cite{norr83}). 
Their photometric and spectroscopic parameters are presented in Table 3. Ten of 
them are CN-strong, with S3839 indices (column 9) larger than 0.33 (our limit 
for $\omega$ Cen), and therefore have been excluded from our metallicity 
calibration (see open triangles in Fig.~3).
Two of the remaining six red giants (I86 and I92) are located on the red
side of the RGB (see Richter et al. \cite{richp}), and their
Str\"omgren colors have to be taken with caution, since a strong reddening
mimics a too low Str\"omgren metallicity. The magnitudes in Table 3 (columns
2-4) have been reddening corrected with $E_{B-V} = 0.42$ (Crocker \cite{croc}),
corresponding to $E_{b-y} = 0.29$
and $E_{m_1} = -0.09$. The identification number of the stars is according
to Arp \& Melbourne (\cite{arpm}). Column 5 gives the metallicity as derived
from our new calibration for Str\"omgren indices (Eq.~(1), Sect.~4). In column
6 and 7, the iron abundances and their references are presented. Values from
Brown \& Wallerstein \cite{brow92} have been shifted by 0.05 dex to match the 
values of Lehnert et al. (\cite{lehn}), according to their different assumption
of the reddening (see also discussion by Anthony-Twarog et al. \cite{anth95}). 
The open circles in column 8 indicate CN-strong stars, filled
circles CN-weak stars. In column 10, the location of the red giant on the
giant branch (red or blue side) according to Richter et al. (\cite{richp}) is
indicated.

In M55, the red giants follow a narrow sequence of a single metallicity in the
$(b-y),m_1$ diagram (see Fig.~2 in Richter et al. \cite{richp}). A linear 
regression to this sequence in the color range $0.5 < (b-y)_0 < 1.1$ represents 
the avarage $(b-y)$ and $m_1$ colors for this metallicity. Since for most
of the red giants no high resolution spectroscopic data are available, a mean
cluster iron abundance of [Fe/H]$= -1.81$ dex (Harris \cite{harr96a}) has been 
adopted. The reddening was assumed to be $E_{B-V} = 0.09$ dex, the mean value
between Harris' list and our independent determination from Str\"omgren colors
(Richter et al. \cite{richp}). This corresponds to $E_{b-y} = 0.06$ and 
$E_{m_1} = -0.02$. For our calibration, four points on the average $(b-y),m_1$
sequence have been chosen for the fit (see Sect.~4 and Fig.~5).

\subsection{The Anthony-Twarog \& Twarog sample}

Anthony-Twarog \& Twarog (\cite{anth98}) compiled a catalog of 360 cool giant
stars, for which they measured Str\"omgren colors and Ca indices, and for
which high-dispersion measurements of iron abundance are available. They give a
metallicity calibration of the $hk$ index (defined as $[(Ca-b)-(b-y)]$) in the
$hk$,$(b-y)$ diagram for cooler, evolved stars.
They homogenized their sample to the
abundance scale of Kraft et al. (\cite{kraf92}), which is consistent with the
Zinn \& West abundance scale (\cite{zinn}). Their $V$ and $(b-y)$ colors 
correspond to the photometric system of Olsen (\cite{olse93}).
The $m_1$ index is tied to the Anthony-Twarog \& Twarog (\cite{anth94a}) system.
In order to be in the same system as our data, the $m_1$ index was corrected
with the transformation given by Olsen (\cite{olse95}).
From their sample we selected all RGB stars with $0.5 < (b-y)_0 < 1.1$ mag
and in the metallicity range of $-2.3 <$ [Fe/H]$_{\rm sp} <
-0.4$ dex. In Table 4 we present the photometric and spectroscopic properties
of the 42 selected stars together with the derived Str\"omgren metallicity
from our new calibration (Eq.~(1)). For the references to the photometric and 
spectroscopic data the
reader is referred to the paper by Anthony-Twarog \& Twarog (\cite{anth98}).

%

\begin{figure}
\psfig{figure=mhilker.f5.ps,height=8.6cm,width=8.6cm
,bbllx=9mm,bblly=65mm,bburx=195mm,bbury=246mm}
\vspace{0.4cm}
\caption{Two-color diagram for the full sample of red giants used for the
Str\"omgren metallicity calibration. Different symbols indicate the
different data sets: filled circles = Anthony-Twarog \& Twarog, asterisks = 
$\omega$ Cen, filled triangles = M22, open circles = avarage data points of 
M55, and open squares = metal-rich data points
from a previous calibration (Grebel \& Richtler \cite{greb92}). Crosses are 
stars excluded from the fit during
the fitting iteration. The bold solid and dashed lines indicate the new 
metallicity calibration between $-$2.5 and 0.0 dex with 4 coefficients
(Eq.~(1)), the thin lines the calibration with 5 coefficients (Eq.~(3))
}
\end{figure}

\section{The Str\"omgren metallicity calibration}

Based on the collected photometric and spectroscopic data, as presented in the
previous sections, our final sample contains 67 red giants, excluding the
CN-strong stars in $\omega$ Cen and M22, and four average data points from M55.
The iron abundances of these
stars range between $-$2.3 and $-$0.5 dex. However, the metal-rich end of our
sample is sparsely populated. Therefore, we connected our calibration
to the more metal-rich sample by Grebel \& Richtler (\cite{greb92}).
We calculated 10 points in the metallicity range between $-$0.5 and 0.0 dex and
in the color range
$0.7 < (b-y)_0 < 1.0$ dex from the previous calibration, and added them to
our list.

\begin{figure}
\psfig{figure=mhilker.f6.ps,height=10.1cm,width=8.6cm
,bbllx=9mm,bblly=65mm,bburx=170mm,bbury=246mm}
\vspace{0.4cm}
\caption{The difference between the Str\"omgren metallicity and spectroscopic 
iron abundances, $\Delta$[Fe/H] =  ([Fe/H]$_{\rm ph} - $[Fe/H]$_{\rm sp}$ is
plotted versus the iron abundances, and the distribution of residuals is shown. 
The symbols are as in Fig.~5: filled circles
= Anthony-Twarog \& Twarog, asterisks = $\omega$ Cen, filled triangles = M22,
open circles = M55, and open squares = metal-rich data points from Grebel \& 
Richtler (\cite{greb92}). Crosses are stars excluded from the fit. The upper
panels show the results for Eq.~(1), the lower ones for Eq.~(3).
The dispersion of the residuals within $\pm 0.25$ dex is 0.11 dex for both 
calibrations, and 0.16 dex when including all stars
}
\end{figure}

Following the calibration by Grebel \& Richtler, a relation of the form
\begin{equation}
{\rm [Fe/H]} = \frac{m_{1,0} + a_1 \cdot (b-y)_0 + a_2}{a_3 \cdot (b-y)_0 + a_4}
\end{equation}
was chosen for the fit. After the first fit with the full sample,
all stars with a deviation in the photometric Str\"omgren metallicity of
more than 0.3 dex compared to the spectroscopic value have been
excluded from the sample (9 stars). Five of these stars are from the $\omega$
Cen sample, and their deviation were already discussed in Sect.~3.2.
The other five deviating stars are from the Anthony-Twarog \& Twarog sample.
All of them have colors bluer than $(b-y) = 0.65$ mag. In this color range
the calibration is less metallicity sensitive, and slightly larger errors
in the $m_1$ index or in [Fe/H]$_{\rm sp}$ can cause large deviations. With the
remaining 58 stars plus the 4 M55 and 10 metal-rich data points the fit was
iterated in such a way that stars with deviations of more than 0.25 dex
were excluded for the next iteration of the fit. Finally, 54 stars and
the 14 other data points remained in the converged fit.
The resulting coefficients are\\
$a_1 = -1.277\pm0.050$ \qquad \qquad $a_2 = \;\;\; 0.331\pm0.035$\\
$a_3 = \;\;\; 0.324\pm0.035$ \qquad \qquad $a_4 = -0.032\pm0.025$

In Fig.~5 the full sample is shown in the two-color diagram together with the
new calibration (bold lines). Fig.~6 (upper panel) shows
$\Delta$[Fe/H] = [Fe/H]$_{\rm ph} - $[Fe/H]$_{\rm sp}$ versus 
[Fe/H]$_{\rm sp}$ for this calibration. The
different symbols in both plots indicate the different data sets and 
stars that have been excluded from the fit.
The dispersion of $\Delta$[Fe/H] around the zero point is 0.16 dex for the
whole sample, or 0.11 dex when accounting only for the stars with residuals
within $\pm 0.25$ dex.
This indicates the average precision of the new Str\"omgren metallicity
calibration for a single giant. Obviously, the precision of a metallicity 
determination is higher in the redder, more metal sensitive part of the color 
range than in the bluer part.

As seen in Fig.~6 (upper panels), most stars with [Fe/H]$_{\rm sp}$ $> -1.0$ 
dex scatter
to negative $\Delta$[Fe/H] values. This might reflect the fact that the 
sensivity to [Fe/H] varies with [Fe/H] in the $(b-y),m_1$ diagram. In order
to account for this we introduced a fifth coefficient and fitted an equation 
of the following form to our data:
\begin{equation}
m_{1,0} = a_1 + a_2\cdot(b-y)_0 + a_3\cdot{\rm [Fe/H]} +
a_4\cdot(b-y)_0\cdot{\rm [Fe/H]} 
\end{equation}
The resulting coefficients are\\
$a_1 = -0.399\pm0.045$ \qquad \qquad $a_2 = \;\;\; 1.354\pm0.057$\\
$a_3 = \;\;\; 0.339\pm0.039$ \qquad \qquad $a_4 = -0.072\pm0.038$\\
$a_5 = -0.011\pm0.009$\\

\noindent
Solving Eq.~(2) for [Fe/H] gives the following form:
\begin{eqnarray}
{\rm [Fe/H]} & = & b_1 + b_2 \cdot (b-y)_0 - [b_3 + b_4 \cdot 
(b-y)_0 + \nonumber \\
 & & b_5 \cdot (b-y)_0^2 + b_6 \cdot m_{1,0}]^{1/2}
\end{eqnarray}
with the coefficients\\
$b_1 = -3.273$ \qquad \qquad $b_2 = 15.409$ \qquad \qquad $b_3 = -25.562$\\
$b_4 = \;22.231$ \qquad \qquad $b_5 = 237.44$ \qquad \qquad $b_6 = -90.909$

The result of Eq.~(3) is shown in Fig.~5 (thin lines) and 
Fig.~6 (lower panels). The metallicity sensitivity as a function of [Fe/H]
is hardly changed, but the slopes of the iso-metallicity 
lines are slightly steeper than in the first calibration.
The scatter at the metal-rich part of the second calibration
seems to be better centred on zero. However, the dispersion is the
same as in the calibration with four coefficients. Since the difference
of both calibrations only is marginal, Eq.~(1) might be used
for simplicity.

In both metallicity calibrations the residuals of the $\omega$ Cen data
scatter to positive values around $\Delta$[Fe/H] $= +0.5\pm0.1$, whereas
the M22 data scatter to negative values around $\Delta$[Fe/H] $= -0.5\pm0.1$. 
An explanation for
these systematic effects is the uncertainty in reddening, especially
in the case of M22. A change of 0.02 mag in $E_{B-V}$ corresponds to a change
of about 0.06 dex in metallicity. Furthermore, in $\omega$ Cen,
some stars with CN band strengths close to the limit of our selection criteria
appear to have a higher metallicity than they actually have (see Fig.~3). 
Since the sample of field giants 
and the M55 data show no systematic deviations, our calibration
seems to be stable against the uncertainties in M22 and $\omega$ Cen. 
In particular, the slope and metallicity of the M55 red giants is very well 
reproduced in the first calibration (see Fig.~6). No trend can be seen for 
$\Delta$[Fe/H] as a function of absolute luminosity of the giants in $\omega$ 
Cen, M22 and M55.

\section{Application to published Str\"omgren data}

\begin{table}
\caption{Determination of metallicities for selected clusters from the 
literature. See text for further explanation and references}
\begin{flushleft}
\begin{tabular}{lrcrr}
\noalign{\smallskip}
\hline
Cluster & \# & $E_{B-V}$ & [Fe/H]$_{\rm ph}$ & 
[Fe/H]$_{\rm lit}$ \\
\noalign{\smallskip}
\hline
\noalign{\smallskip}
NGC~6334 & 11 & 0.09 & $-0.23\pm0.12$ & $-$0.16...0.23 \\
         & 11 & 0.14 & $0.01\pm0.14$ & \\[2pt]
NGC~3680 &  5 & 0.04 & $-0.14\pm0.06$ & $-$0.16...0.10 \\
         &  5 & 0.09 & $0.11\pm0.07$ & \\[2pt]
NGC~2395 & 25 & 0.10 & $-0.69\pm0.29$ & $-$0.70...$-$0.36 \\
         & 24 & 0.21 & $-0.26\pm0.28$ & \\[2pt]
NGC~6397 & 29 & 0.16 & $-2.20\pm0.19$ & $-$2.21...$-$1.85 \\
         &  6 & 0.20 & $-2.17\pm0.35$ & \\[2pt]
M55      & 63 & 0.07 & $-1.88\pm0.11$ & $-$1.95...$-$1.65 \\
         & 48 & 0.14 & $-1.61\pm0.12$ & \\[2pt]
M22      & 63 & 0.32 & $-1.91\pm0.18$ & $-$1.75...$-$1.56 \\
         & 44 & 0.42 & $-1.52\pm0.19$ & \\
\noalign{\smallskip}
\hline
\end{tabular}
\end{flushleft}
\end{table}

Only few Galactic globular and open clusters have been studied in the 
Str\"omgren system in the last two decades, most of them by Anthony-Twarog, 
Twarog and collaborators. Only recently Grundahl et al. (1999) started a
programme of deep $uvby$ CCD imaging of several globular clusters in the 
Milky Way. In order to compare our calibration (Eq.~(1))
with published data we determined average metallicities of red giants for the
open clusters NGC 6334 (Anthony-Twarog \& Twarog \cite{anth87}), NGC 3680 
(Anthony-Twarog et al. \cite{anth89}), and NGC 2395
(Anthony-Twarog et al. \cite{anth94b}), and the globular clusters NGC 6397
(Anthony-Twarog \& Twarog \cite{anth92}) M55, and M22 (Richter et al. 
\cite{richp}).
The determined Str\"omgren metallicity depends on the adopted 
reddening. In Table 5 we present the results for our calibration (column 4) 
in the range of published reddenings (column 3) in comparison to otherwise 
determined iron abundances (column 5, see Harris \cite{harr96a}, Friel 
\cite{frie}, and the Anthony-Twarog et al. papers for references). 
Column 2 gives the number of stars involved in the metallicity determination.
Our results agree well with the published iron abundances. Also
the iso-metallicity lines of the red giants in the $(b-y),m_1$ diagram
agree very well with the slopes of the new calibration for all clusters.
NGC 2395 has a high intrinsic
dispersion similar to M22 and $\omega$ Cen. The high dispersion for NGC 6397 
is due to the fact that all giants
have $(b-y)$ colors bluer than 0.55 mag, a very metal insensitive regime 
for Str\"omgren metallicities.

\section{Summary}

Red giants in the globular clusters $\omega$ Centauri, M55,
and M22 together with field giants from Anthony-Twarog \& Twarog 
(\cite{anth98}) have been used to revise the Grebel \& Richtler (\cite{greb92})
metallicity calibration of the Str\"omgren $(b-y),m_1$ diagram. 
For all giants in $\omega$ Cen and M22, accurate and homogeneous iron abundances
from high resolution spectroscopy are available in the literature. M55 has a
well determined average iron abundance value.
In total, 62 CN-weak giants have been used. CN-rich stars have been excluded,
since their $m_1$ value mimics a too high iron abundance in the $(b-y),m_1$ 
diagram.
In order to cover a wide metallicity range, $-2.0 <$[Fe/H]$< 0.0$ dex, our
new calibration is connected to a previous calibration by 
Grebel \& Richtler (\cite{greb92}) around solar metallicities. 
In the color range  $0.5 < (b-y)
< 1.1$ mag, for which our calibration is valid, the loci of equal iron 
abundances lie on straight lines.

We emphasize that it was possible to find a uniform metallicity calibration
in the indicated parameter range that seems to have no obvious dependencies
on luminosity within the errors. In particular, the new calibration seems to 
be valid for globular cluster as well as for field giants.
No variation of metallicity sensitivity with metallicity in the 
$(b-y),m_1$ diagram has been found.

On average, the precision of a metallicity determination with our new 
calibration for a single giant is in the order of 0.11 dex.
Average abundances of giants within a cluster and relative abundances between
clusters can be determined with much higher precision, depending on the
number of red giants per cluster sample.
The application of the new calibration to independent samples of red giants
with published Str\"omgren photometry agrees very well with otherwise 
determined abundances.

For the red giants in $\omega$ Cen and M22, the influence of CN-strong stars
on the metallicity calibration has been studied. For Str\"omgren metallicities
higher than $-$1.0 dex, CN-weak stars cannot be distinguished in the 
$(b-y),m_1$ diagram
from stars with lower iron abundances but higher CN band strengths.
However, the difference between the Str\"omgren metallicity of CN-rich stars
and their spectroscopically determined iron abundance is correlated to the CN 
band strengths. This might be used to determine the iron
abundance of a red giant if its Str\"omgren colors and one of the CN indices
are known, or alternatively to detect CN-rich stars when their iron
abundances are known.

%

\acknowledgements
This research was supported through `Pro- yecto FONDECYT 3980032'.
I thank Tom Richtler and Boris Dirsch for interesting discussions, and
the referee for very helpful comments that improved the paper.

\enddocument